\documentclass{ws-procs9x6}

\def\Pbar{\overset{\rule{2mm}{.2mm}}{P}}
\def\Mbar{\overset{\rule{2mm}{.2mm}}{M}}
\def\Gbar{\overset{\rule{2mm}{.2mm}}{\Gamma}}
\def\Kbar{\overset{\rule{2mm}{.2mm}}{K}}
\def\ud{\mathrm{d}}
\def\mD{\mathcal{D}}

\begin{document}

\title{Helicity amplitudes and electromagnetic decays of strange
  baryon resonances}

\author{ \underline{T. Van
    Cauteren}\footnote{e-mail:\uppercase{T}im.\uppercase{V}an\uppercase{C}auteren@\uppercase{UG}ent.be} and J. Ryckebusch}

\address{Ghent University, B-9000 Gent, Belgium}

\author{B. C. Metsch and H.-R. Petry}

\address{Helmholtz-Institut f\"ur Strahlen- und Kernphysik \\
Bonn University, D-53115 Bonn, Germany}

\maketitle

\abstracts{
We present results for the helicity amplitudes of the lowest-lying
hyperon resonances $Y^*$, computed within the framework of the Bonn
constituent-quark model, which is based on the Bethe-Salpeter
approach~\cite{merten1,tim1,tim2}. The seven parameters entering the
model are fitted against the best known baryon
masses~\cite{loeringphd}. Accordingly, the results for the helicity
amplitudes are genuine predictions. Some hyperon resonances are seen
to couple more strongly to a virtual photon with finite $Q^2$ than to
a real photon. Other $Y^*$'s, such as the $S_{01}(1670)$ $\Lambda$
resonance or the $S_{11}(1620)$ $\Sigma$ resonance, have large
electromagnetic decay widths and couple very strongly to real
photons. The negatively-charged and neutral members of a $\Sigma^*$
triplet may couple only moderately to the $\Sigma(1193)$, while the
positively-charged member of the same $\Sigma^*$ triplet displays a
relatively large coupling to the $\Sigma^{+}(1193)$ state. This
illustrates the necessity of investigating all isospin channels in
order to obtain a complete picture of the hyperon spectrum.}

\section{Introduction}\label{sec:intro}

The implementation of the electromagnetic (EM) couplings to a hadron
resonance in isobar models constitutes one of its major sources of
uncertainty. This is particularly the case for models including
hyperon resonances, for which little experimental information is
available. These $Y^*$'s can play an important role in the background
of kaon electroproduction processes, photo- and electroproduction of
the elusive $\Theta(1540)$ resonance, and radiative kaon capture
reactions. In this work, the Bonn constituent-quark (CQ) model, based
on the Lorentz-covariant Bethe-Salpeter approach, is used to compute
the electromagnetic helicity amplitudes (HA's) of the lowest-lying
hyperon resonances. The results may be used in isobar models where a
$\gamma^{(*)} Y Y^*$ coupling gets introduced.

\section{Helicity Amplitudes in the Bethe-Salpeter Approach}\label{sec:model}

The expression for the current matrix elements in the framework of the
Bonn CQ model can be found in Refs.~\refcite{merten1,tim1,tim2}
and~\refcite{mertenphd}. In the rest frame of the incoming baryon
resonance, the EM transition matrix element reads~:
\begin{multline}
\langle \, \Pbar \, | \, j^\mu \, | \, \Mbar^* \, \rangle \simeq -3 \iint
\frac{\ud^4 \left[ \frac{1}{2} \left( p_1 - p_2 \right) \right]} {(2
\pi )^4} \frac{\ud^4 \left[ \frac{1}{3} \left( p_1 + p_2 - 2p_3\right)
\right]} {(2 \pi )^4} \\
\times \, \Gbar^{\Lambda}_{\Pbar} \; S^1_F (p_1) \otimes S^2_F (p_2) \otimes
\left[ S^3_F (p_3 + q) \, \hat{q} \gamma^\mu S^3_F (p_3) \right] \;
\Gamma^{\Lambda}_{\Mbar^*} \; ,
\label{eq:CME}
\end{multline}
where $\Gamma$ and $\Gbar$ are the amputated BS amplitude and its
adjoint of the incoming and outgoing baryon respectively, and $S^i_F$
is the $i$'th CQ propagator. In the above matrix element, the operator
$\hat{q} \gamma^\mu$ describes how the photon couples to a pointlike
CQ with charge $\hat{q}$.

The electromagnetic properties of baryon resonances are usually
presented in terms of helicity amplitudes, which are functions of the
squared fourmomentum of the photon. The HA's are directly proportional
to the current matrix element with the appropriate spin projections
for incoming and outgoing hyperon~:
\begin{subequations}
\begin{multline}
A_{1/2} \left( B^* \rightarrow B \right) \, = \, \mD \, \left\langle B,
\Pbar, \frac{1}{2} \right| j^1 (0) + i \, j^2 (0) \left| B^*, \Pbar^*,
-\frac{1}{2} \right\rangle \; , \label{eq:HA_def_a12}
\end{multline}
\begin{multline}
A_{3/2} \left( B^* \rightarrow B \right) \, = \, \mD \, \left\langle B,
\Pbar, -\frac{1}{2} \right| j^1 (0) + i \, j^2 (0) \left| B^*,
\Pbar^*, -\frac{3}{2} \right\rangle \; , \label{eq:HA_def_a32}
\end{multline}
\begin{equation}
C_{1/2} \left( B^* \rightarrow B \right) \, = \, \mD \, \left\langle
B, \Pbar, \frac{1}{2} \right| j^0 (0) \left| B^*, \Pbar^*, \frac{1}{2}
\right\rangle \; , \label{eq:HA_def_c12}
\end{equation}
\label{eq:HA_def}
\end{subequations}
with $\mD = \sqrt{\frac {\pi \alpha} {2 M ({M^*}^2 - M^2)}}$, where
$\alpha$ is the fine-structure constant and $M^*$ ($M$) the mass of
the incoming (outgoing) hyperon.

\section{Results and Conclusions}\label{sec:results}

\begin{figure}
\centerline{\epsfxsize=0.9\textwidth\epsfbox{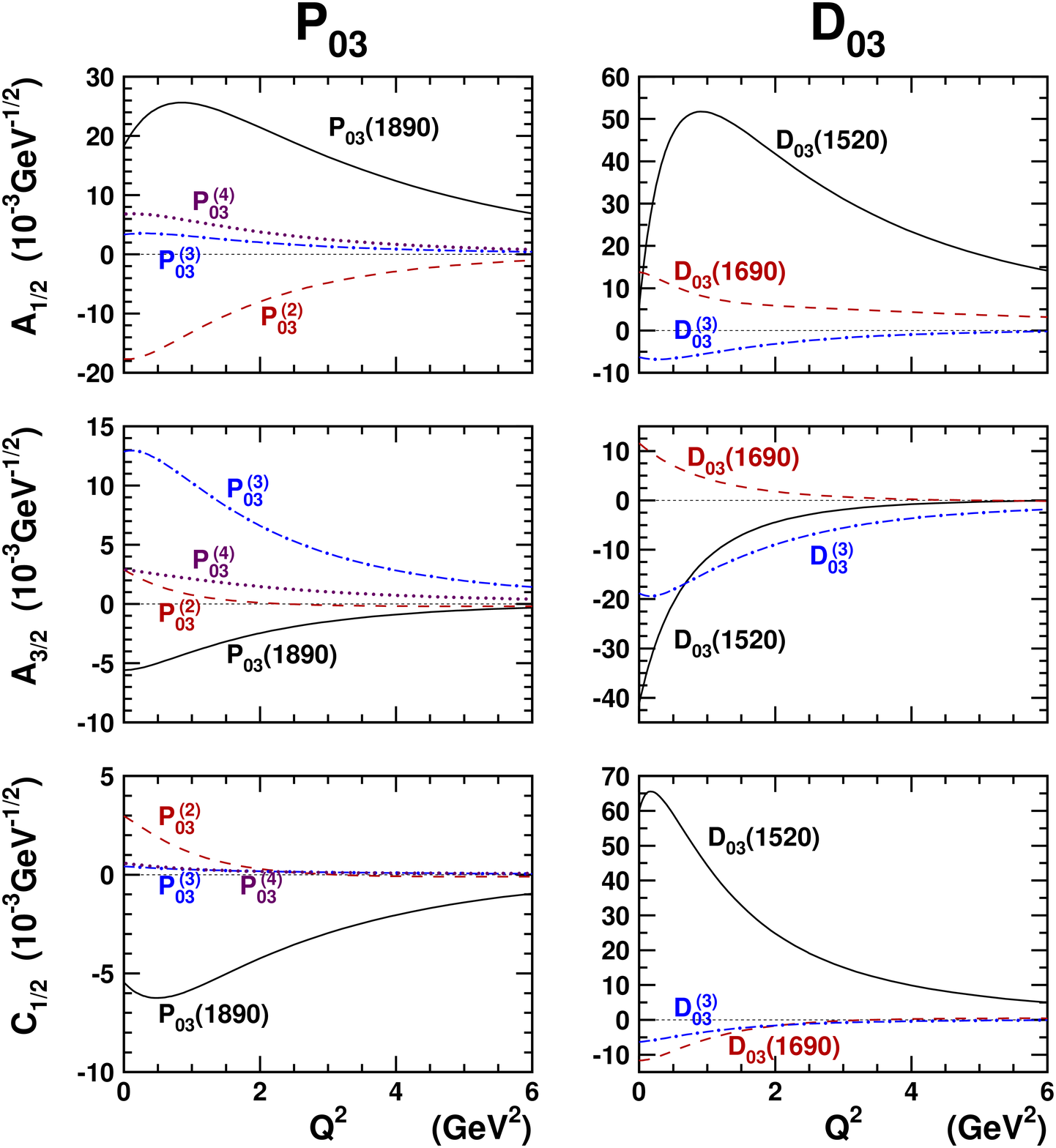}}
\caption{The $Q^2$ dependence for the $\Lambda^* + \gamma^* \to
    \Lambda$ decays for spin $J=3/2$ resonances~: left (right) panels
    show the results for the positive (negative) parity $\Lambda^*$
    resonances. \label{fig:spin_3_2_iso_0}}
\end{figure}

In Fig.~\ref{fig:spin_3_2_iso_0} we present our predictions for the
HA's of the lowest-lying spin $J=3/2$ $\Lambda$ resonances, decaying
to the $\Lambda(1116)$. One notices that the $A_{1/2}$ and $C_{1/2}$
of the lightest resonance with either parity show a maximum at finite
$Q^2$. This feature also occurs in some of the other HA's we have
computed within the Bonn CQ model~\cite{tim2}. As a consequence,
$u$-channel contributions to $p(e,e'K)Y$ processes, can be expected to
exhibit a peculiar $Q^2$ dependence.

In Table~\ref{tab:dec_wid}, the EM decay widths of the hyperon
resonances with masses around $1660$~MeV are given. This is the energy
region investigated by the Crystal Ball Collaboration in Brookhaven
for the $\Kbar^- p \to \gamma Y^0$ processes~\cite{cb_col}. Our
results indicate that the $S_{01}(1670)$ and $D_{03}(1690)$
$\Lambda$-resonances could be important if $Y^0 = \Sigma^0(1193)$. If
on the other hand $Y^0 = \Lambda(1116)$, the $P_{11}(1660)$,
$S_{11}(1620)$ and $D_{13}(1670)$ $\Sigma$-resonances seem to be more
appropriate candidates for governing the reaction dynamics.

\begin{table}
\tbl{Electromagnetic decay widths of the hyperon resonances around
  $1660$~MeV to the different ground-state hyperons in units of MeV.}
{\footnotesize
\begin{tabular}{|c|cccc|}
\hline
{} &{} &{} &{} &{}\\[-1.5ex]
$Y^*$ & $\Gamma_{Y^* \to \Lambda(1116)}$ & $\Gamma_{{Y^*}^0 \to
  \Sigma^0(1193)}$ & $\Gamma_{{Y^*}^+ \to \Sigma^+(1193)}$ &
$\Gamma_{{Y^*}^- \to \Sigma^-(1116)}$\\[1ex]
\hline
{} &{} &{} &{} &{}\\[-1.5ex]
$S_{01}(1670)$ & $0.159 \times 10^{-3}$ & $3.827$ & --- & ---\\[1ex]
$D_{03}(1690)$ & $0.0815$ & $1.049$ & --- & ---\\[1ex]
\hline
{} &{} &{} &{} &{}\\[-1.5ex]
$P_{11}(1660)$ & $0.451$ & $0.0578$ & $0.733$ & $0.141$\\[1ex]
$S_{11}(1620)$ & $1.551$ & $0.688$ & $5.955$ & $0.613$\\[1ex]
$D_{13}(1670)$ & $1.457$ & $0.0214$ & $0.440$ & $0.184$\\[1ex]
\hline
\end{tabular}\label{tab:dec_wid} }
\vspace{8pt}
\end{table}

Another distinct feature illustrated by Table~\ref{tab:dec_wid} is the
dependence of the EM decay widths of the $\Sigma$ resonances on the
isospin-3 component. For positively-charged $\Sigma^*$'s, the reported
decay widths can be an order of magnitude larger than for the
negatively-charged or neutral members of the isospin triplet. These
resonances could therefore contribute to the background of the
$p(\gamma,K^0)\Sigma^+$ process significantly, while being marginal
for the $p(\gamma,K^+)\Sigma^0$ reaction.

\end{document}